          [2001/04/25 1.1 (PWD)]
\documentclass[a4paper,twocolumn]{esapub} % European paper
\usepackage{natbib,psfig}

\title{On the stability of self-gravitating protoplanetary discs}
\author[1]{W.K.M. Rice}
\author[2,3]{P.J. Armitage}
\author[1]{I.A. Bonnell}
\author[4]{M.R. Bate}
\affil[1]{School of Physics and Astronomy, University of St Andrews,
North Haugh, St Andrews, KY16 9SS, Scotland}
\affil[2]{JILA, Campus Box 440, University of Colorado, Boulder CO 80309-0440, USA}
\affil[3]{Department of Astrophysical and Planetary Sciences, University of
Colorado, Boulder CO 80309-0391, USA}
\affil[4]{School of Physics, University of Exeter, Stocker Road, Exeter
EX4 4QL, UK}

\begin{document}

\maketitle

\begin{abstract}
It has already been shown, using a local model, that accretion discs with cooling times $t_{\rm cool}
\le 3 \Omega^{-1}$ fragment into gravitationally bound objects, while those with cooling times $t_{\rm cool}
> 3 \Omega^{-1}$ evolve into a quasi-steady state. We present results of three-dimensional simulations
that test if the local result still holds globally. We find that the fragmentation boundary is close to that
determined using the local model, but that fragmentation may occur for longer cooling times when the disc is more
massive or when the mass is distributed in such a way as to make a particular region of the disc more susceptible to
the growth of the gravitational instability. These results have significant implications for the formation of of gaseous
planets in protoplanetary discs and also for the redistribution of angular momentum which could be driven by the
presence of relatively massive, bound objects within the disc.
\end{abstract}

\section{Introduction}
The discovery of the first extra-solar planet \citep{mayor95} and the subsequent discovery
of many ($> 100$) additional extra-solar planets \citep{marcy00} has 
enhanced the interest in the
formation of planets and the evolution of protoplanetary discs. The observation of the first 
transiting planet \citep{henry00,charbonneau00}, giving an estimate of the planet radius, and the relatively large
masses ($> 0.1 M_{\rm Jupiter}$) of all the currently detected extra-solar planets, has led to the
view that they are all giant gaseous planets. The most widely studied, and accepted, 
formation mechanism for gaseous planets is core accretion \citep{lissauer93} in which planetesimals grow
by direct collision to form a core which, when sifficiently massive ($m \sim 10 m_{\rm earth}$),
then accretes an envelope of gas from the disc. Models \citep{pollack96}, however, suggest that
the timescale for planet formation via this mechanism may be longer than the lifetime of 
most protoplanetary accretion disc \citep{haisch01}. This
has led to a renewed interest in the possibility that gas giant planets may have formed directly
via a disc instability \citep{kuiper51,boss98,boss00,mayer02}.

A protoplanetary disc that is sufficiently massive and cool may form gravitationally bound 
gaseous objects via the gravitational instability. This mechanism is extremely rapid and 
differs from core accretion in that a rocky core is not required. The possibility that this 
mechanism may play a role in gaseous planet formation has been enhanced by recent models of the interiors
of gaseous planets \citep{guillot97, guillot99} which suggest that Jupiter could have a relatively small core with
$m_{\rm core} < 10 m_{\rm earth}$. A Keplerian accretion disc with sound speed $c_s$, surface density
$\Sigma$, and epicyclic frequency $\kappa$ will become gravitationally unstable if the \cite{toomre64}
$Q$ parameter 
\begin{equation}
Q=\frac{c_s \kappa}{\pi G \Sigma}
\end{equation}
is of order unity. A gravitationally unstable disc can either fragment into one or more gravitationally
bound objects, or it can evolve into a quasi-steady state in which gravitational instabilities lead
to the outward transport of angular momentum. The exact outcome depends on the rate at which the
disc heats up (through the dissipation of turbulence and gravitational instabilities) and the
rate at which the disc cools. It has been suggested \citep{goldreich65} that a feedback loop may
exist. When $Q$ is large the disc is stable and cooling dominates, driving the disc towards
instability. When $Q$ becomes sufficiently small, heating through viscous dissipation
dominates, and the disc is returned to a state of marginal stability. In this way $Q$ is maintained
as a value of $\sim 1$.

It has, however, been shown using a local model \citep{gammie01} that a quasi-stable state can only
be maintained if the cooling time $t_{\rm cool} > 3 \Omega^{-1}$ where $\Omega$ is the local angular
frequency. For shorter cooling times the disc fragments. This is consistent with \cite{pickett98,
pickett00} that `almost isothermal' conditions are necessary for fragmentation, and defines a robust
lower limit to the critical cooling time below which fragmentation occurs. It has been suggested,
however, that self-gravitating discs require a strictly global treatment \citep{balbus99}, and
while global effects are highly unlikely to stabilize a locally unstable disc, they could well
allow fragmentation within discs that would be locally stable. This has led to an interest
in global simulations of self-gravitating protoplanetary discs \citep{rice03} and we discuss, in
this paper, results of a number of these simulations.  

\section{Numerical Simulations}

\subsection{Smoothed particle hydrodynamics}

The three dimensional, global simulations presented here were performed using
smoothed particle hydrodynamics (SPH), a Lagrangian hydrodynamics code \citep{benz90,monaghan92}.
The central star is modelled as a point mass that may accrete gas particles if they
approach within a predefined sink radius, while the gaseous disc is modelled
using 250000 SPH particles. A tree is used to determine neighbours and to calculate
gravitational forces between gas particles and between gas particles and point masses 
\citep{benz90}. If in the simulation the disc starts to fragment, producing high density regions, the
code can in principle continue. The high density regions do, however, slow the code
down significantly. To continue simulating a fragmenting disc, high density regions that are 
gravitationally bound are converted into point masses \citep{bate95} which may continue to accrete disc gas.
Since the likely number of point masses is quite small, compared to the number of gas
particles, the gravitational force between point masses is computed directly. An additional
saving in computational time is also made by using individual particle time-steps \citep{bate95,navarro93}.
The time-steps for each particle is limited by the Courant condition and by a force condition
\citep{monaghan92}.

\subsection{Initial conditions}

We consider a system comprising a central star, modelled as a point mass with mass $M_*$, surrounded
by a gaseous circumstellar disc with mass $M_{\rm disc}$. We performed a number
of simulations with disc masses of $M_{\rm disc} = 0.1 M_*$ and $M_{\rm disc} = 0.25 M_*$. 
The disc temperature is taken 
to have an initial radial profile of $T \propto r^{-0.5}$ \citep{yorke99} 
and in most of the simulations the Toomre $Q$ parameter is assumed to be initially constant
and to have a value of $2$. A stable accretion disc where self-gravity leads to a steady
outward transport of angular momentum should have a near constant $Q$ throughout. A constant
$Q$ together with Equation 1 gives a surface density profile of $\Sigma \propto r^{-7/4}$, and hydrostatic
equilibrium then gives a central density profile of $\rho \propto r^{-3}$. We did, however, perform  
a single simulation with $M_{\rm disc} = 0.1 M_*$ and with $\Sigma \propto r^{-1}$. In this case, $T \propto
r^{-0.5}$ gives a $Q$ that is
not constant but that decreases with increasing radial distance. The temperature was chosen such that the
initial minimum value of $Q$ was 1.5.

The disc is modelled using 250000 SPH particles, which are initially randomly distributed such as
to give the specified density profile between inner and outer radii of $r_{\rm in}$ and $r_{\rm out}$.
Since the calculations are scale free, we take $M_* = 1$, $r_{\rm in} = 1$, and $r_{\rm out} = 25$. If
we were to assume a physical mass scale of $1$ M$_\odot$ and a length scale of $1$ au, the central
star would have a mass of $1$ M$_\odot$, the circumstellar disc would have a mass of $0.1$ M$_\odot$ 
or $0.25$ M$_\odot$ and would extend from $1$ to $25$ au, and 1 yr would equal $2\pi$ code units.

\subsection{Cooling}
The main aim of this work is to test if the results obtained by \cite{gammie01} still hold
globally. Consequently we use the same approach in these simulations as was used in the local
model. We use an adiabatic equation of state, with adiabatic index $\gamma = 5/3$, and allow the
gas to heat due to both PdV work and dissipation. Cooling is implemented by adding a simple
cooling term to the energy equation. Specifically, for a particle with internal energy per unit
mass $u_i$, 
\begin{equation}
\frac{{\rm d} u_i}{{\rm d} t} = -\frac{u_i}{t_{\rm cool}}
\end{equation}
where, as in \cite{gammie01}, $t_{\rm cool}$ is given by $\beta \Omega^{-1}$ with the value
of $\beta$ varied for each run.

Although the above cooling time is essentially chosen to compare the local model results
with results using a global model, it can also be related (at least approximately) to
the real physics of an accretion disc. For an optically thick disc in equilibrium, the
cooling time is given by the ratio of the thermal energy per unit area to the radiative losses
per unit area. It can be shown \citep{pringle81} that in such a viscous accretion disc, the
cooling time is
\begin{equation}
t_{\rm cool} = \frac{4}{9 \gamma(\gamma-1)}\frac{1}{\alpha \Omega}
\label{pringle}
\end{equation}
where $\Omega$ is the angular frequency in the disc, and $\alpha$ is the \cite{shakura73}
viscosity parameter.

\section{Results}

\subsection{$M_{\rm disc} = 0.1 M_*$}
We use a three dimensional, global model to consider how cooling times of $t_{\rm cool} = 
5 \Omega^{-1}$, and $t_{\rm cool} = 3 \Omega^{-1}$ affect the gravitational stability of a
protoplanetary accretion disc with $M_{\rm disc} = 0.1 M_*$ and with $\Sigma \propto r^{-7/4}$.
This choice of surface density profile, together with $T \propto r^{-0.5}$, gives a Toomre $Q$
parameter that is intially constant throughout the disc. 

\begin{figure}
\centerline{\psfig{file=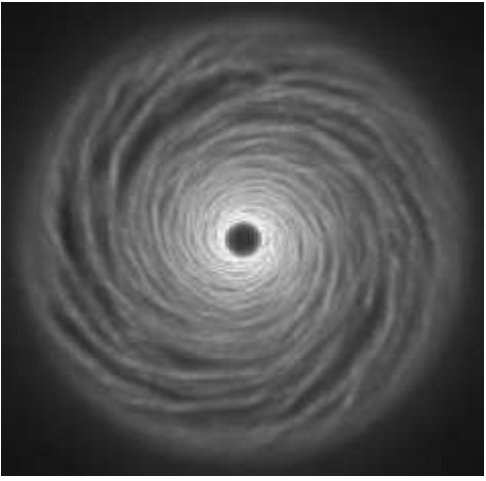,width=0.8\linewidth}}
\caption{Equatorial density structure for a disc with $M_{\rm disc} = 0.1 M_*$, 
$\Sigma \propto r^{-7/4}$ and with a cooling time of $t_{\rm cool} = 5 \Omega^{-1}$. The density
is highly structured with the instability present at all radii. The density has, however, not
increased significantly, and the disc is in a quasi-steady state with the heating through
the dissipation of the instability balancing the imposed cooling.\label{tc5disc}}
\end{figure}

\begin{figure}
\centerline{\psfig{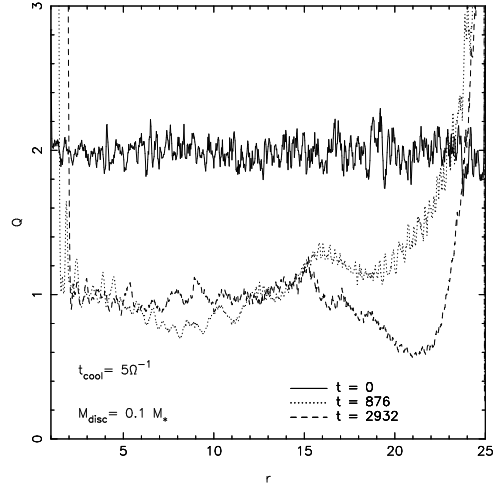}}
\caption{Toomre $Q$ parameter at the beginning ($t=0$), one third of the way ($t=876$), and at
the end ($t=2932$) of the simulation in which $M_{\rm disc} = 0.1 M_*$, $\Sigma \propto r^{-7/4}$,
and $t_{\rm cool} = 5 \Omega^{-1}$. At the end of the simulation $Q$ is of order unity between
radii of 1 and 15.\label{tc5Q}}
\end{figure}

Figure \ref{tc5disc} shows the final equatorial density structure of the $t_{\rm cool} = 5 \Omega^{-1}$
simulation. The central star (not shown) is located in the middle of the figure, and the $x$ and $y$
axes both run from $-25$ to $25$. The disc is highly structured and the instability exists at all 
radii. However, at no point in the disc has the density increased significantly, and no fragmentation 
has taken place. Figure \ref{tc5Q} shows the Toomre $Q$ parameter for the same simulation at three
different times during the simulation. At the beginning of the simulation ($t=0$) $Q$ has an almost
constant value of $2$. At the end of the simulation ($t=2932$) the value of $Q$ is almost unity between
radii of $1$ and $15$. Comparing $Q$ at $t=876$ and $t=2932$ shows that the cooling initially
reduces $Q$ to a value below 1, causing the instability to grow, heating the disc and returning $Q$
to a value of order unity. The disc has clearly settled into a quasi-steady state in which the imposed 
cooling is balanced by heating through the dissipation of the gravitational instability.

\begin{figure}
\centerline{\psfig{file=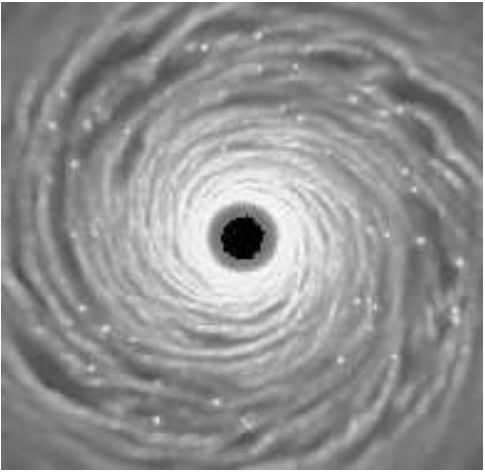,width=0.8\linewidth}}
\caption{Equatorial density structure of the inner regions of a disc with $M_{\rm disc} = 0.1 M_*$,
$\Sigma \propto r^{-7/4}$, and with a cooling time of $t_{\rm cool} = 3 \Omega^{-1}$. The disc
is highly structured and is starting to fragment into high density, gravitationally bound
clumps.\label{tc3disc}}
\end{figure}

Figure \ref{tc3disc} shows the equatorial disc structure for a simulation with the same
disc mass and surface density profile as in Figure \ref{tc5disc}, but with a cooling time of
$t_{\rm cool} = 3 \Omega^{-1}$. This simulation was only
run for 504 time units and hence we show only the inner 8 radii of the simulation. The disc is
again highly structured, but in this case the disc has started to fragment into gravitationally bound
clumps, indicated by the bright dots in Figure \ref{tc3disc}. The density in the clumps is
4 - 5 orders of magnitude greater than the initial disc density, and to reach the time shown
in Figure \ref{tc3disc} many of the clumps have been converted into point masses \citep{bate95}. 

\begin{figure}
\centerline{\psfig{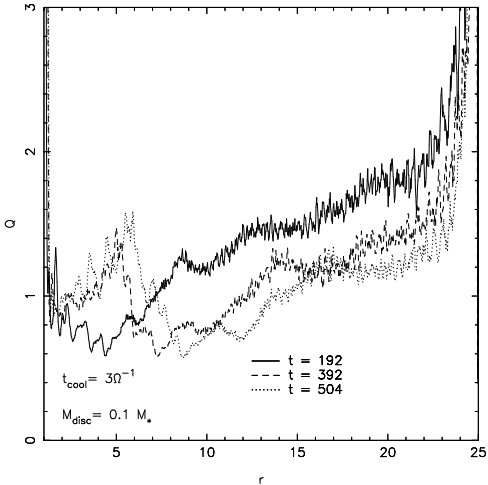}}
\caption{Toomre $Q$ parameter at times of $t = 192$, $t = 392$, and $t = 504$ for $M_{\rm disc}
= 0.1 M_*$, $\Sigma \propto r^{-7/4}$, and $t_{\rm cool} = 3 \Omega^{-1}$. \label{tc3Q}}
\end{figure}

Figure \ref{tc3Q} shows the Toomre $Q$ parameter at three different times during the above
simulation. At all three times shown, there is a region which has $Q$ significantly less than $1$.
In the regions where $Q$ is less than unity, the instability grows rapidly and produces gravitationally bound
fragments. Once formed, these fragments tidally heat the disc, increasing $Q$ to a value
greater than unity. Once $Q > 1$, fragmentation ceases and the disc, at that radius, becomes
gravitationally stable. There is, however, likely to be a significant amount of angular momentum
transport driven by tidal interactions between the gravitationally bound fragments and the 
gaseous disc \citep{larson02}.

\subsection{$M_{\rm disc} = 0.25 M_*$}
To study how disc mass may influence the global nature of the gravitational
instability we consider a disc with a mass of $M_{\rm disc} = 0.25 M_*$. As in
the previous simulations, the surface density is taken to have a radial profile
of $\Sigma \propto r^{-7/4}$ which, together with $T \propto r^{-0.5}$, gives an
intially constant Toomre $Q$ parameter. Although most T Tauri discs have masses
considerably less than $0.25 M_*$ \citep{beckwith90}, there a few with such masses, and this simulation
may also apply to an earlier stage of the star formation process when discs are
expected to be more massive.

\begin{figure}
\centerline{\psfig{file=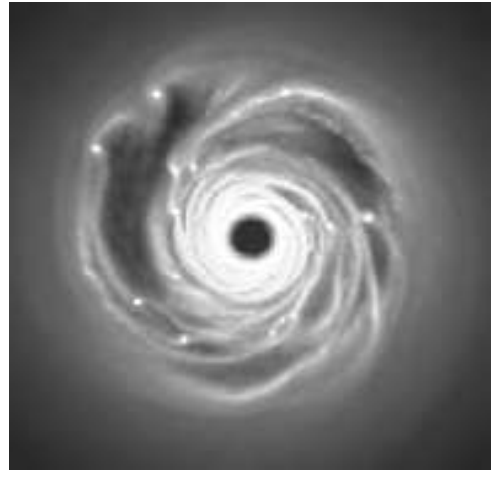,width=0.8\linewidth}}
\caption{Equatorial density structure of a disc with $M_{\rm disc} = 0.25 M_*$, 
$\Sigma \propto r^{-7/4}$, and $t_{cool} = 5 \Omega^{-1}$. There are clear signs
of fragmentation with the most massive fragments being gravitationally boud.
\label{tc5Md025disc}}
\end{figure}

Figure \ref{tc5Md025disc} shows the final equatorial density structure of the 
above simulation. The disc is highly structured and the nature
of the spirals, compared to the equivalent simulation with $M_{\rm disc} = 0.1 M_*$, 
is consistent with the increased disc mass \citep{nelson98}. Unlike the equivalent
lower mass simulation (see Figure \ref{tc5disc}), there are a number of high density
clumps present in the disc. The standard routine for checking if these clumps
are bound initially found them to be unbound.  This routine, however, only considers
the nearest $\sim 50$ SPH neighbours. By increasing this to the nearest $\sim 250$
SPH neighbours, the densest clump was found to be just gravitationally bound. This
would suggest that in more massive discs, global effects could act to make the disc
more unstable, allowing gravitationally bound fragments to grow for cooling times
greater than that obtained using a local model. This could have significant
implications for the formation of planets, or binary companions \citep{adams89},
reasonably early in the star formation process.

\subsection{$M_{\rm disc} = 0.1 M_*$, $\Sigma \propto r^{-1}$}

\begin{figure}
\centerline{\psfig{file=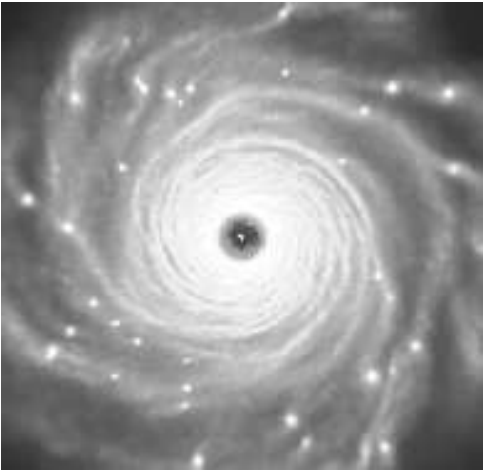,width=0.8\linewidth}}
\caption{Equatorial density structure of a disc with $M_{\rm disc} = 0.1 M_*$, 
$\Sigma \propto r^{-1}$, and $t_{cool} = 5 \Omega^{-1}$. There are clear signs
of fragmentation suggesting that the fragmentation boundary may depend both on the
mass of the disc and on how the mass is distributed.
\label{tc5sigr-1}}
\end{figure}

All of the previous simulations were performed assuming $\Sigma \propto r^{-7/4}$,
and $T \propto r^{-0.5}$, giving an initially constant Toomre $Q$ parameter. We have
performed a single simulation with $M_{\rm disc} = 0.1 M_*$, $T \propto r^{-0.5}$, and
$\Sigma \propto r^{-1}$. With these parameters, the $Q$ value decreases with
increasing radius. We therefore normalised our temperature such that $Q = 1.5$ at
$r = 25$. Figure \ref{tc5sigr-1} shows the final equatorial density structure of
a simulation with the above disc parameters, and with an imposed cooling time
of $t_{\rm cool} = 5 \Omega^{-1}$. The disc is again highly structured and
unlike the simulation with the same disc mass and cooling time, but with the
steeper surface density profile, there is clear evidence of fragmentation in the
outer regions of the disc. The shallower
surface density profile ($\Sigma \propto r^{-1}$) means that, 
compared to the steeper surface
density profile ($\Sigma \propto r^{-7/4}$), there is more mass at large radii. This suggests
that the global nature of the instability may depend both on the mass of the disc, and
on how the mass is distributed within the disc. 

\section{Conclusions}

It has been shown using a local model \citep{gammie01} that a disc will fragment
for cooling times $t_{\rm cool} \le 3 \Omega^{-1}$ and will settle into a quasi-steady
state for cooling times $t_{\rm cool} > 3 \Omega^{-1}$. We present here results from
three-dimensional simulations which test if the local results still holds globally \citep{rice03}. 
We impose the same cooling function as in \cite{gammie01} which, although fairly simplistic, can
also be justified physically \citep{pringle81}. 

For a disc mass of $M_{\rm disc} = 0.1 M_*$, a cooling time of $t_{\rm cool} = 5 \Omega^{-1}$, and
an initially constant $Q$, the disc settles into a quasi-steady state in which $Q$ is of order unity
over a large region of the disc. Decreasing the cooling time to $t_{\rm cool} = 3 \Omega^{-1}$ causes
the disc to rapidly become unstable and produces numerous gravitationally bound clumps. For this disc mass
and surface density profile ($\Sigma \propto r^{-7/4}$), the fragmentation boundary therefore seems to agree with
that obtained using a local model.

Increasing the disc mass to $M_{\rm disc} = 0.25 M_*$ and keeping all other parameters unchanged, however, 
results in fragmentation for a cooling time of $t_{\rm cool} = 5 \Omega^{-1}$. A similar result was 
obtained for a simulation in which the disc mass was kept at $M_{\rm disc} = 0.1 M_*$ but in which the surface density
profile was changed from $\Sigma \propto r^{-7/4}$ to $\Sigma \propto r^{-1}$. These results suggest that
global effects may become significant for massive disc or for discs in which the mass is distributed in such
a way as to make a particular region of the disc susceptible to the growth of the gravitational instability. 
For T Tauri discs, which generally have masses $M_{\rm disc} < 0.1 M_*$ \citep{beckwith90}, the fragmentation
boundary is therefore likely to be close to that determined using the local model. Earlier in the star formation process,
when disc may be more massive, fragmentation may occur for longer cooling times than that predicted by
the local model results \citep{gammie01}.

For quiescent T Tauri discs, the \cite{shakura73} $\alpha$ is conventially estimated to be of the
order of $10^{-2}$ or smaller \citep{hartmann98,bell94}. The simple estimate quoted earlier (see
Eq. \ref{pringle}) would suggest that such discs are comfortably stable. \cite{boss02}, however, has
shown using an approximate treatment of disc heating and cooling, that there may be periods when the
cooling time is comparable to the orbital period. Our results suggest that if the disc is fairly massive,
such short cooling times may open a window of opportunity for the formation of substellar objects, probably
in the form of a multiple system 
\citep{armitage99}. This has important implifications for the formation of giant gaseous planets in
protoplanetary discs and may also play a role in the rapid transfer of angular momentum \citep{larson02}.

\section*{Acknowledgments}

The simulations reported in this paper made use of the UK Astrophysical Fluids Facility (UKAFF).
WKMR acknowledges support from a PPARC standard grant.

% The following bibliography was produced with
%   \bibliographystyle{aa}
%   \bibliography{esapub}
% The results are inserted directly here to simplify
% the demonstration.

\end{document}